\documentclass{PoS}

\newcommand{\be}{\begin{equation}}
\newcommand{\ee}{\end{equation}}
\newcommand{\bea}{\begin{eqnarray}}
\newcommand{\eea}{\end{eqnarray}}

\newcommand{\dslash}{Dslash }

\usepackage{listings}
\usepackage{color}
\usepackage[usenames,dvipsnames]{xcolor}
\usepackage{amsfonts}
\usepackage{subfigure}

\usepackage{url}
\definecolor{backcolor}{rgb}{0.95,0.95,0.92}
\definecolor{codegreen}{rgb}{0,0.6,0}
\definecolor{codegray}{rgb}{0.5,0.5,0.5}
\definecolor{codepurple}{rgb}{0.58,0,0.82}

\lstdefinestyle{mystyle}{
    backgroundcolor=\color{backcolor},   
    keywordstyle=\color{codepurple},
    commentstyle=\color{Emerald},
    stringstyle=\color{magenta},
    numberstyle=\tiny\color{codegray},
    basicstyle={\footnotesize\ttfamily},
    breaklines=true,                 
    captionpos=b,                    
    keepspaces=true,                 
    showspaces=false,                
    showstringspaces=false,
    showtabs=false,                  
    tabsize=2
}

\title{Optimizing the domain wall fermion Dirac operator using the R-Stream source-to-source compiler
}

\ShortTitle{Optimizing the domain wall fermion Dirac operator using R-Stream}
\author{\speaker{Meifeng Lin}$^a$,  Eric Papenhausen$^b$, M. Harper Langston$^c$, Benoit Meister$^c$,  Muthu Baskaran$^c$, Taku Izubuchi$^{d,e}$ and Chulwoo Jung$^d$  \\
        $^a$Computational Science Center, Brookhaven National Laboratory, Upton, NY 11973, USA\\
        $^b$Department of Computer Science, State University of New York, Stony Brook, NY 11794, USA \\
        $^c$Reservoir Labs Inc., New York, NY 10012, USA \\
        $^d$Physics Department, Brookhaven National Laboratory, Upton, NY 11973, USA \\
        $^e$RIKEN-BNL Research Center, Brookhaven National Laboratory, Upton, NY 11973, USA \\
        E-mail: \email{mlin@bnl.gov}, \email{epapenhausen@cs.stonybrook.edu}, \email{langston@reservoir.com}, \email{meister@reservoir.com}, \email{baskaran@reservoir.com},  \email{izubuchi@bnl.gov}, \email{chulwoo@bnl.gov}}
 
\abstract{The application of the Dirac operator on a spinor field, the \dslash operation, is the most computation-intensive part of the lattice QCD simulations. It is often the key kernel to optimize to achieve maximum performance on various platforms. Here we report on a project to optimize the domain wall fermion Dirac operator in Columbia Physics System (CPS) using the R-Stream source-to-source compiler. Our initial target platform is the Intel PC clusters.  We discuss the optimization strategies involved before and after the automatic code generation with R-Stream and present some preliminary benchmark results.}

\FullConference{The 33rd International Symposium on Lattice Field Theory\\
		14 -18 July 2015\\
		Kobe International Conference Center, Kobe, Japan}

\begin{document}

\section{Introduction}

Lattice QCD (LQCD) community has traditionally produced very efficient home-grown software and is continuing to do so. With the advent of new hardware architectures, significant efforts are required to optimize the LQCD software for new systems. One way to deal with this is to rewrite the software to be \emph{future-proof}, which inevitably needs significant initial investments to develop a software suite that is portable, flexible, adaptable and, most importantly, performant for different architectures. Another way, perhaps complementary to the first approach, is to develop or use existing high-quality automatic \emph{code generators} that are capable of producing efficient codes for a target architecture from generic, high-level, user codes. The latter will help to port an existing application to a new architecture quickly, and ideally with good performance. 

Numerical LQCD computations are dominated by the application of the Dirac operator matrix $D$ on a fermion field vector $\psi$, the so-called \dslash operation. It is often the key kernel to optimize for maximum performance, as the \dslash operation typically accounts for more than 90\% of the execution time in LQCD simulations. In this work we explore the feasibility and efficiency of using the R-Stream source-to-source compiler being developed by Reservoir Labs Inc to optimize the \dslash code in the Columbia Physics System (CPS)~\cite{cps}. This report is organized as following. After a brief introduction to R-Stream in Section~\ref{sec:rstream}, we present the details of the application of R-Stream to the Wilson \dslash and the Domain Wall Fermion (DWF) \dslash in Section~\ref{sec:implementation}. Section~\ref{sec:simd} discusses SSE and AVX SIMD vectorizations on the R-Stream generated code. In Section~\ref{sec:performance} we show the performance benchmarks for the single-node Wilson and DWF Dslash. And we conclude in Section~\ref{sec:conclusions}. 

\section{The R-Stream Source-to-Source Compiler}\label{sec:rstream}

R-Stream~\cite{rstream} is a source-to-source compiler that generates target-specific optimized source codes based on the polyhedral mapping for computer programs~\cite{polyhedral,memory-alloc}. It takes a serial C code as input, generates the dependence graph using the polyhedral model, and can perform optimizations ranging from loop-level parallelism, tiling to memory management. The outputs of R-Stream may be C code threaded with OpenMP to run on CPUs, or CUDA C for NVIDIA GPUs. These outputs can then be processed by a low-level compiler such as \texttt{icc}, \texttt{gcc} or \texttt{nvcc}. The advantage of using a high-level code generator such as R-Stream is that the source code generated can be tuned and further optimized before sending it to a low-level compiler, allowing user-in-the-loop optimizations. For more details of the types of optimizations R-Stream can perform and the process involved, readers can refer to ~\cite{rstream, MemoryReuse, VISSOFT15}.

\section{Application of R-Stream to \dslash Operators in CPS}\label{sec:implementation}

Our first application of R-Stream is the Wilson \dslash operator, which is the kernel operator in many lattice fermion formulations, including the domain wall fermion formulation used in many of the lattice simulations performed by the RBC and UKQCD collaborations. The Wilson quark Dirac matrix can be written as 
\be
M = (N_d + m_q) - \frac{1}{2} D,
\label{eq:action_Wilson}
\ee
where $N_d$ is the space-time dimension and $m_q$ is the input quark mass. The Wilson \dslash operator, $D$, in four space-time dimensions is defined as 
\bea
D_{\alpha\beta}^{ij}(x,y) &=& \sum_{\mu=1}^4 \left [ (1-\gamma_\mu)_{\alpha\beta} {U_\mu}^{ij}(x) \delta_{x+\hat\mu,y} + 
(1+\gamma_\mu)_{\alpha\beta} {U^\dagger_\mu}^{ij}(x+\hat\mu)\delta_{x-\hat\mu,y} \right ],
\label{eq:wilson-dslash}
\eea
where $x$ and $y$ are the coordinates of the lattice sites, $\alpha, \beta$ are spin indices, and $i, j$ are color indices. $U_\mu(x)$ is the gluon field variable and is an SU(3) matrix. In CPS, the \emph{complex} fermion fields are represented by one-dimensional arrays with size $L_X L_Y L_Z L_T\times SPINS \times COLORS\times 2$ for the un-preconditioned Dirac operator, where $L_X, L_Y, L_Z$ and $L_T$ are the numbers of lattice sites in the $x, y, z$ and $t$ directions, respectively. $SPINS$ and $COLORS$ are the numbers of spin and color degrees of freedom, typically 4 and 3 respectively. With even-odd preconditioning, the array sizes are cut in half. 

The domain wall fermion (DWF) Dirac matrix $M_{x,s; x^\prime, s^\prime}$ is defined as 
\begin{equation} M_{x,s; x^\prime, s^\prime} = \delta_{s,s^\prime}
M^\parallel_{x,x^\prime} + \delta_{x,x^\prime} M^\bot_{s,s^\prime}, \label{eq:dwf}
\end{equation}
where $s, s'$ label the extra fifth dimension, and the 4D DWF Dslash is defined as 
\begin{eqnarray} M^\parallel_{x,x^\prime} & =& - {1\over 2} \sum_{\mu=1}^4 \left[
(1-\gamma_\mu) U_{x,\mu} \delta_{x+\hat\mu,x^\prime} + (1+\gamma_\mu)
U^\dagger_{x^\prime, \mu} \delta_{x-\hat\mu,x^\prime} \right]  +
 (4 - M_5)\delta_{x,x^\prime}. \label{eq:D_parallel} 
\end{eqnarray}
Note that Eq.(\ref{eq:D_parallel}) is just the Wilson Dirac
operator in Eq.(\ref{eq:action_Wilson}) with a negative mass $ m_q\equiv -M_5$ and is independent of $s$. The hopping term in the fifth dimension is defined as 
\begin{eqnarray} M^\bot_{s,s^\prime} &=& - {1\over
2}\Big[(1-\gamma_5)\delta_{s+1,s^\prime} + (1+\gamma_5)\delta_{s-1,s^\prime} -
2\delta_{s,s^\prime}\Big] \nonumber\\ &+& {m_f\over 2}\Big[(1-\gamma_5)
\delta_{s, L_s-1} \delta_{0, s^\prime} +
(1+\gamma_5)\delta_{s,0}\delta_{L_s-1,s^\prime}\Big].  \label{D_perp}
\end{eqnarray} 
As the 4D DWF Dslash dominates DWF calculations, we first focused on its optimization. 

The input we gave to R-Stream was the \texttt{noarch} version of the Wilson \dslash operator in CPS, which is an unoptimized serial C code. 
For the polyhedral mapping to work, the array accesses have to be \emph{affine} functions of the outer loop indices so that the dependence can be constructed as a system of linear equations. 
The array access in CPS breaks affinity, as the array index for the lattice site $(x,y,z,t)$ has to be calculated as products of the array indices. To overcome this, we cast the linear arrays as multi-dimensional pointers in C, which gives us the performance of linear arrays but gives R-Stream the readability of multi-dimensional arrays.  

Another change we had to make to the input \dslash  code was to remove the modulo operations when the lattice boundaries are involved. For example, in the $x$ direction, the \dslash operation for the site $x=LX-1$ requires the access to the fermion vector at site $x=0$, which is done by \texttt{x \% LX}. Similarly, access to $x=LX-1$ is needed when \dslash operates on $x=0$. To preserve affinity, we padded each boundary with the value at the opposite boundary. The new fermion arrays have a size $LX+2$ in the $x$ direction, with \texttt{psi[0]} containing the value of $x=LX-1$, and \texttt{psi[LX+1]} containing the value of $x=0$. Similarly for the other directions. For the gauge field, since only the neighbor in the forward direction is needed, we only need to introduce padding for one side of the boundaries, resulting in a new array length of $L+1$ in each of the space-time directions. The downside of the data padding is that the memory footprint is increased substantially, especially when local lattice volume is small. However, when the memory increases the most (small local volume) is also when the total memory footprint is small, even though the percentage increase may be large. So having padding does not affect the problem size we are able to simulate greatly. 

With the above two modifications, R-Stream was able to analyze the input code and generate outputs based on the user guidance. Details of how to use R-Stream and how the user can affect the optimization strategies are beyond the scope of these proceedings, and we refer the readers to Refs.~\cite{rstream, VISSOFT15} for further information. The R-Stream generated code included automatic loop  tiling and unrolling. 
While the performance of the R-Stream output was a bit better than the input unimproved serial code, it was much inferior to a version of a hand-optimized CPS code which has SSE intrinsics. To fully exploit the performance offered by the modern computer architectures, it is imperative for us to take advantage of their SIMD capabilities. We thus further optimized the code using both the Intel SSE and AVX intrinsics, which we will discuss in the next section. 

\section{SIMD Implementation}\label{sec:simd}
\subsection{SSE}
The Intel SSE instruction set extension allows to perform two double-precision or four single-precision floating point operations at each clock cycle. It requires the data in such operations to be vectorized. As discussed in Section~\ref{sec:implementation}, we use the multi-dimensional arrays for the fermion vectors. For the Wilson \dslash in 4D, it goes as 
\begin{lstlisting}[language=C]
double psi[LT][LZ][LY][LX][4][3][2];
\end{lstlisting}
where the real/imaginary index runs the fastest. For double precision, this data structure is already vectorized in a way suitable for the SSE instruction. While using this vectorization does not require a complete data structure transformation, the cross term in the complex multiplication requires permutation of the complex vector, which may affect the performance. We used this implementation for the performance benchmark shown in Section~\ref{sec:performance}, as our final goal is to implement the AVX instruction for the DWF Dslash operator. 

\subsection{AVX}
The AVX instruction set extension can perform four double-precision or eight single-precision floating point operations at one clock cycle, provided that the data are vectorized accordingly. A data structure transformation is required to implement the AVX vectorization. There are various ways to do this. For the Wilson Dslash, we chose to vectorize using the four spinor degrees of freedom. The new data layout becomes 
\begin{lstlisting}[language=C]
double psi[LT][LZ][LY][LX][3][2][4];
\end{lstlisting}
For DWF, we tried different data layouts, including one similar to the Wilson Dslash. But the best performance was obtained when the data layout vectorized along the fifth dimension:
\begin{lstlisting}[language=C]
double psi[LT][LZ][LY][LX][3][2][4][LS];
\end{lstlisting}
as the 4D DWF \dslash in Eq.(~\ref{eq:dwf}) is inherently data parallel in the fifth dimension. The disadvantage of this data layout is, it requires \texttt{LS} to be multiples of 4 (8) when double (single)  precision is used in the computation. A more flexible data structure is discussed in~\cite{grid}, which we will look into in the future.

\section{Performance}\label{sec:performance}
\subsection{Compile Environment and Test Platforms}
For the performance benchmarks reported in this section, double precision arithmetics were used. OpenMP was used for threading, while SSE or AVX intrinsics were used for SIMD. In the tests shown below, we used the GNU C compiler \texttt{gcc} version 5.1.0. Similar results were obtained with \texttt{gcc} version 4.9.2. We ran the tests on three clusters: (i) The \texttt{pi0} cluster at Fermilab with dual-socket Intel ``Sandy Bridge" Xeon CPU E5-2650 v2 at 2.60GHz, referred to as \texttt{SNB-pi0}. (ii) The \texttt{hpc1} cluster at Brookhaven National Lab with dual-socket Intel ``Sandy Bridge" Xeon CPU E5-2670 at 2.60 GHz, referred to as \texttt{SNB-hpc1}. (iii) The \texttt{lired} cluster at Stony Brook University with dual-socket Intel ``Haswell" Xeon CPU E5-2690 v3 at 2.60 GHz, referred to as \texttt{HSW-lired}.  Both \texttt{SNB-pi0} and \texttt{SNB-hpc1} support AVX extension, while \texttt{HSW-lired} supports AVX2. 

\subsection{Single-Node Wilson Dslash}
\begin{figure}[tp]
\centering
\vspace{-.5cm}

\subfigure[ ] {
\includegraphics[width=0.46\textwidth,clip]{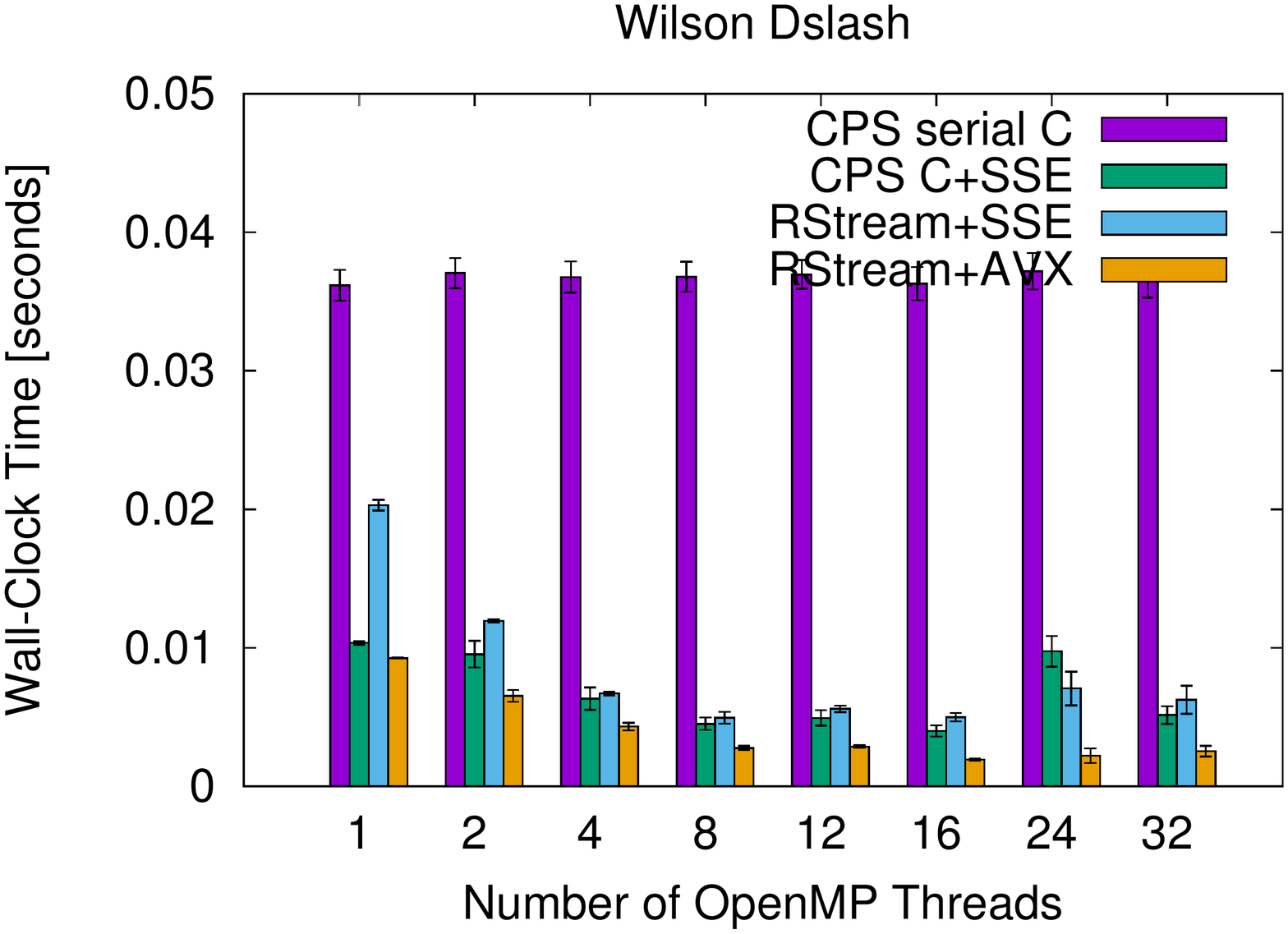}
\label{fig:wilson-all}
}
\hfill
\subfigure[] {
\includegraphics[width=0.46\textwidth,clip]{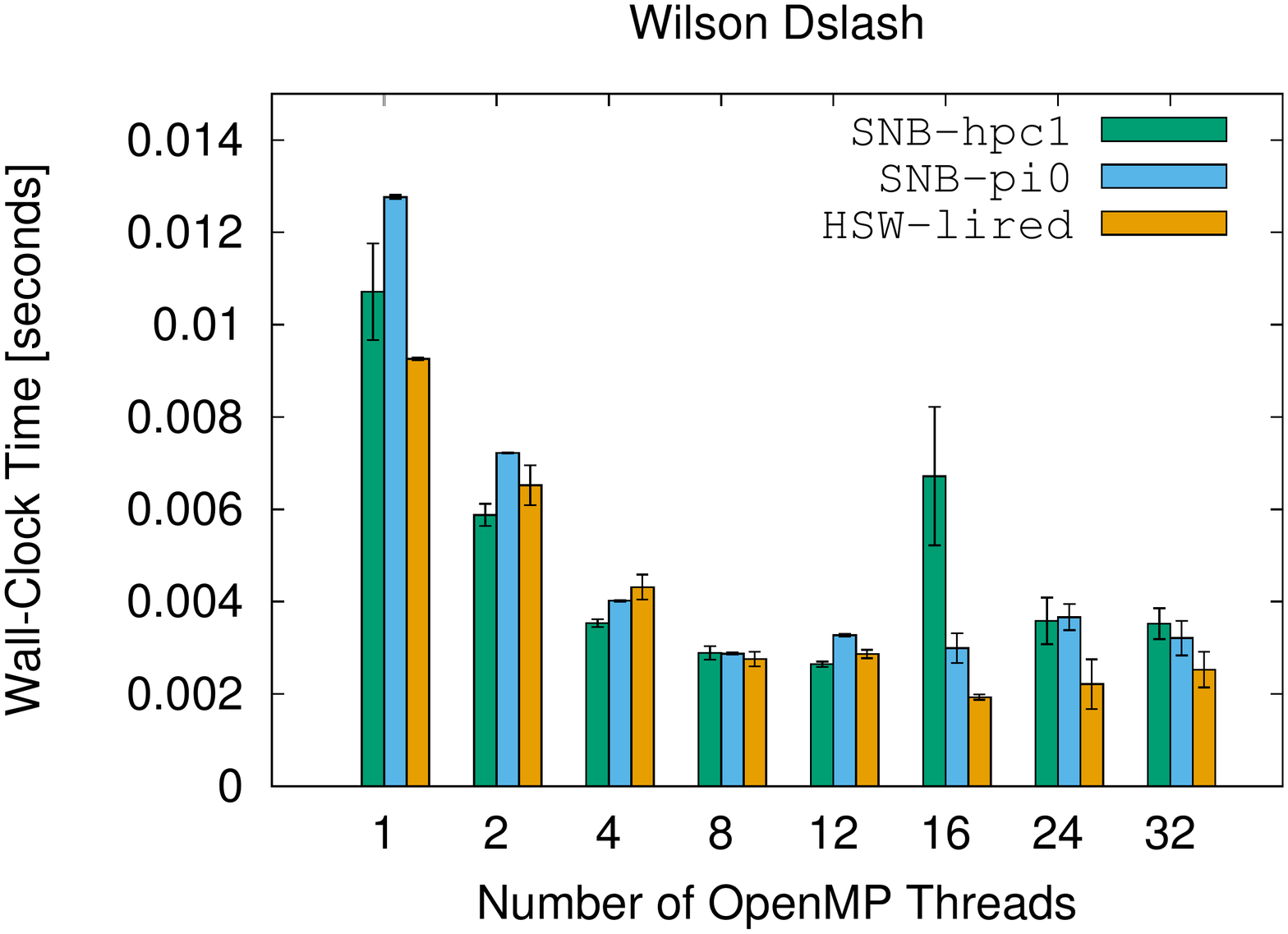}
\label{fig:wilson-machine}
}
\caption{Single-node performance for the Wilson Dslash on a $16^4$ lattice. Figure~\protect\ref{fig:wilson-all} shows the comparison of wall-clock time for four implementations of the even-odd preconditioned Wilson Dslash on \texttt{HSW-lired}. Figure~\protect\ref{fig:wilson-machine} shows the comparison of wall-clock time for RStream+AVX implementation on three different machines. Timing is averaged over 10 runs, and the error bars show its variance.\label{fig:wilson}}
\end{figure}

Figure~\ref{fig:wilson-all} shows the comparison of wall-clock time for the even-odd preconditioned Wilson \dslash in several implementations on a $16^4$ lattice using the \texttt{HSW-lired} cluster. ``CPS serial C" refers to the \texttt{noarch} implementation in CPS. ``CPS C+SSE" refers to the original hand-optimized version with SSE intrinsics. ``RStream+SSE" refers to the R-Stream generated code with SSE intrinsics and ``RStream+AVX" refers to the version with AVX intrinsics. The best performance was achieved with 16 threads for the parallelized versions, with the AVX implementation performing twice as well as the CPS SSE version, at about 22 GFlops per node. While the one-thread performance with AVX was quite good, at 4.6 GFlops, the performance suffered from poor scaling of our OpenMP implementation, which we expect to improve with further investigation. 

We also show the performance for the RStream+AVX implementation on three different machines in Figure~\ref{fig:wilson-machine}. The performances were quite similar. While the Haswell processors support fused multiply-add (FMA) that are available in AVX2, at the time of our implementation we did not have access to this type of CPUs and thus did not include AVX2 instructions. It is not surprising that our code performs similarly on these three machines.  It is also worth noting that the total number of cores on one node is 16 for the Sandy Bridge clusters and 24 for the Haswell cluster. When the number of OpenMP threads is equal to the number of cores available, performance drops slightly, possibly due to the way the threads are scheduled. 
 
\subsection{Single-Node 4D DWF Dslash}
\begin{figure}[tp]
\centering
\vspace{-.5cm}
\subfigure[ ] {
\includegraphics[width=0.46\textwidth,clip]{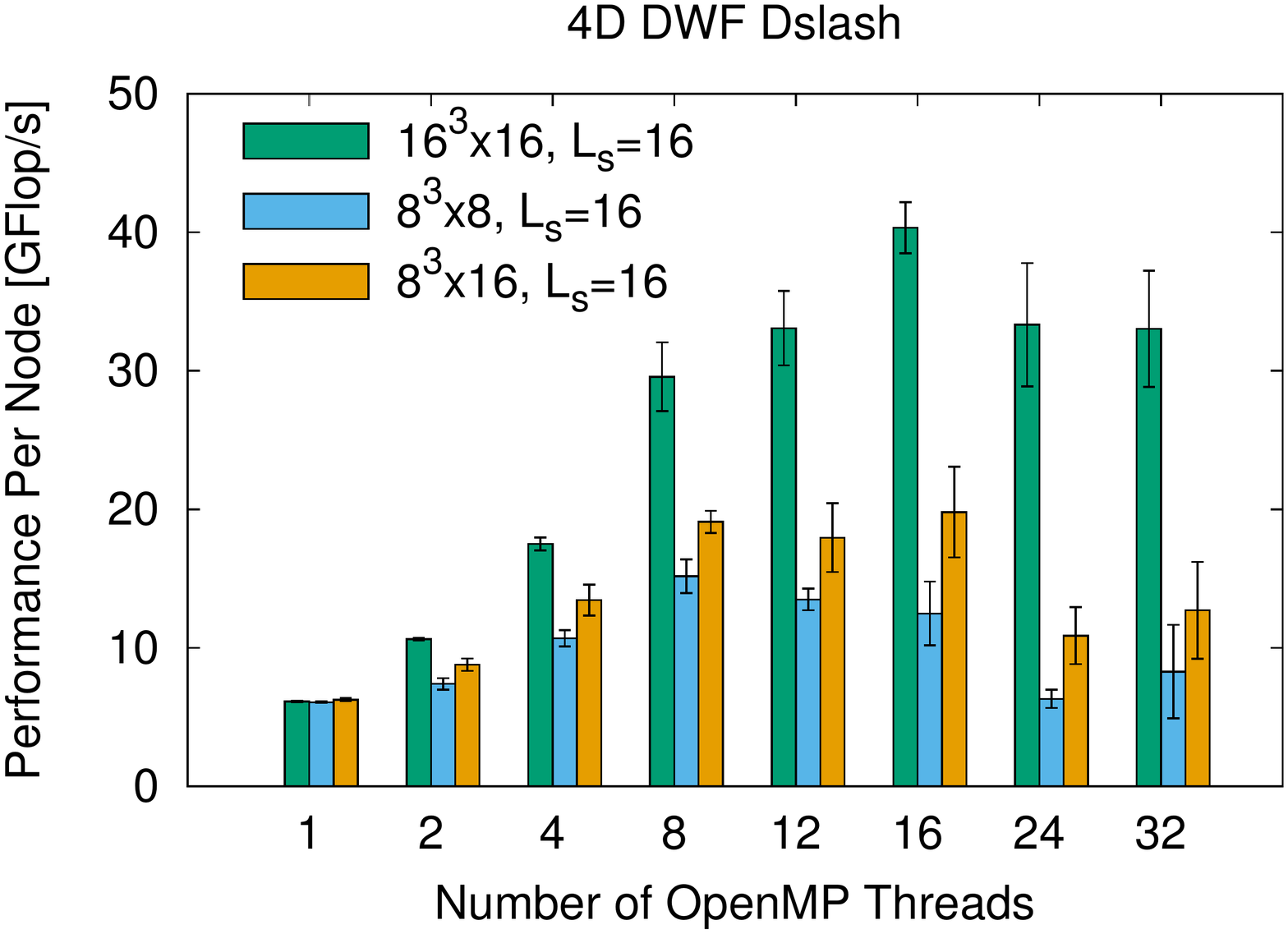}
\label{fig:dwf-volume}
}
\hfill
\subfigure[] {
\includegraphics[width=0.46\textwidth,clip]{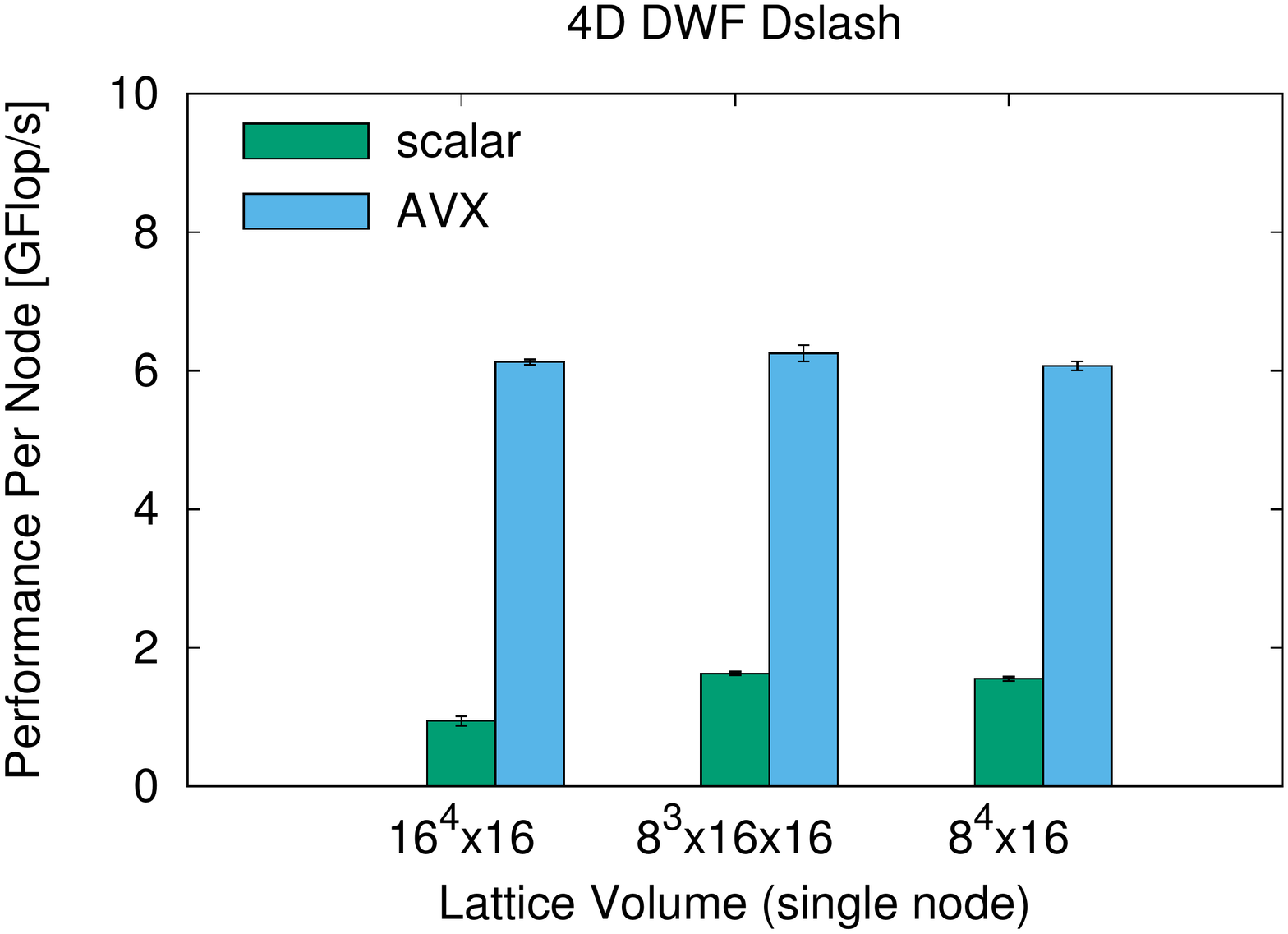}
\label{fig:dwf-vec}
}
\caption{Single-node performance  for the RStream+AVX implementation of the 4D DWF \dslash on \texttt{HSW-lired}. Figure~\protect\ref{fig:dwf-volume} shows the performance with different OpenMP threads. Figure~\protect\ref{fig:wilson-machine} shows the effect of AVX vectorization on different lattice volumes with one OpenMP thread. Timing is averaged over 10 runs, and the error bars show its variance.  }\label{fig:dwf}
\end{figure}

For DWF, the dominating computation is the 4D DWF Dslash as given in Eq.(\ref{eq:D_parallel}). We show the performance of the 4D DWF Dslash with ``RStream+AVX" implementation in Figure~\ref{fig:dwf}. The tests were run on the \texttt{HSW-lired} cluster. Figure~\ref{fig:dwf-volume} shows the performance on one node with different lattice volumes. Since we are interested in scaling up to a large number of nodes, we also tested the small node-volumes of $8^4$ and $8^3\times16$. The fifth dimension length $L_s$ was fixed to 16. The best performance was achieved with the node-volume of $16^4$ and with 16 OpenMP threads at about 40 GFlops. We also show the effect of vectorization in Figure~\ref{fig:dwf-vec}, where we compare the performance of the scalar CPS \texttt{noarch} version with  the R-Stream generated code further vectorized with AVX. As expected, at least a factor of 4 speedup was achieved. 

\section{Summary and Conclusions}\label{sec:conclusions}
We have presented our experience with using the R-Stream source-to-source, polyhedral-model-based, compiler to optimize the \dslash operator in CPS for both the Wilson and DWF fermions. With some modifications to the input sequential C code, we were able to use R-Stream to analyze and generate C code with loop reorganizations that made it easier for us to parallelize with OpenMP for the single-node execution. With AVX SIMD instructions, we were able to achieve more than a factor of 16 speedup on a single node compared to the input serial code for the Wilson \dslash and more than a factor of  40 for the 4D DWF \dslash. The performance is twice that of the manually optimized CPS code with SSE intrinsics, suggesting that such high-level code generators have the potential to give performance on par with hand-written codes. More tests with different lattice volumes and different compile environments are ongoing. We are also working on a single-precision version that should offer even greater performance. Multi-node version with MPI communications is in progress. The performance for the full version with conjugate gradient, MPI communications and single precision arithmetics will be presented in a future publication. 

\section*{Acknowledgments}
This material is based upon work supported by the U.S. Department of Energy, Office of Science, Office of Nuclear Physics under Award Number DE-SC0009678. M.L., T.I. and C.J.  are supported in part by the U.S. Department of Energy, Office of Science under Contract Number DE-SC0012704 through which Brookhaven National Laboratory is operated. T.I. is also supported by Grants-in-Aid for Scientific Research \#26400261. We gratefully acknowledge the use of computing resources on the USQCD cluster at Fermilab, HPC Code Center cluster at Brookhaven National Laboratory and the LI-red cluster at Stony Brook University for testing and benchmarking.

\bibliographystyle{JHEP-2}
\bibliography{refs}
\end{document}